\documentclass[twocolumn]{aastex631}

\usepackage{color}
\usepackage{url}
\usepackage{hyperref}
\usepackage{xspace}
\usepackage{float}
\usepackage{soul}
\usepackage{bm}
\usepackage{mathtools}
\usepackage{enumerate}

\newcommand {\ixpe}{{IXPE}\xspace}
\newcommand{\rmd}{{\rm d}}
\renewcommand{\vec}[1]{\mbox{\boldmath ${#1}$}}
\newcommand{\taut}{\tau_0}
\newcommand{\sigmat}{\sigma_{\rm T}}

\submitjournal{ApJ Letters}

\shorttitle{X-ray polarization of Cyg X-1}
\shortauthors{Poutanen et al.}

\begin{document}

\title{Polarized X-rays from windy accretion in Cygnus X-1}

 \author[0000-0002-0983-0049]{Juri Poutanen}
\affiliation{Department of Physics and Astronomy, FI-20014 University of Turku,  Finland}

\correspondingauthor{Juri Poutanen}
\email{juri.poutanen@utu.fi}

\author[0000-0002-5767-7253]{Alexandra Veledina}
\affiliation{Department of Physics and Astronomy, FI-20014 University of Turku,  Finland}
\affiliation{Nordita, KTH Royal Institute of Technology and Stockholm University, Hannes Alfv\'{e}ns v\"{a}g 12, SE-106\,91 Stockholm, Sweden}

\author[0000-0001-5660-3175]{Andrei M. Beloborodov}
\affiliation{Physics Department and Columbia Astrophysics Laboratory, Columbia University, 538 West 120th Street, New York, NY 10027, USA}
\affiliation{Max Planck Institute for Astrophysics, Karl-Schwarzschild-Str. 1, D-85741 Garching, Germany} 




\begin{abstract}
Recent X-ray polarimetric data on the prototypical black hole X-ray binary Cyg~X-1 from the Imaging X-ray Polarimetry Explorer present tight constraints on accretion geometry in the hard spectral state.
Contrary to general expectations of a low, $\lesssim1\%$ polarization degree (PD), the observed average PD was found to be a factor of 4 higher.
Aligned with the jet position angle on the sky, the observed polarization favors geometry of the X-ray emission region stretched normally to the jet in the accretion disk plane.
The high PD is, however, difficult to reconcile with the low orbital inclination of the binary $i\approx30\degr$.
We suggest that this puzzle can be explained if the  emitting plasma is outflowing with a mildly relativistic velocity $\gtrsim0.4\,c$. 
Our radiative transfer simulations show that Comptonization in the outflowing medium elongated in the plane of the disk radiates X-rays with the degree and direction of polarization consistent with observations at $i\approx30\degr$.
\end{abstract}



\section{Introduction} 
\label{sec:intro}

Accreting black holes (BH) in X-ray  binaries display different spectral states,  ``hard'' and ``soft'', distinguished by the spectral shape and the variability properties \citep{ZG04,RM06}.
In the soft state, dominant contribution to the X-ray flux comes from a thermally looking spectrum peaking at $\sim$1~keV energies, which is commonly attributed to the multi-temperature blackbody accretion disk \citep{1973A&A....24..337S,NT73}. 
An additional power-law-like tail is likely produced by inverse Compton scattering by relativistic electrons in a corona \citep{PC98,Gierlinski1999,Zdziarski01,McConnell02}. 
In the hard state, the spectrum is power-law-like in the standard X-ray band and shows a cutoff at $\sim$100 keV, which is interpreted as a signature of thermal Comptonization in a hot electron medium. 
The geometry of this medium and the source of soft seed photons is a matter of debates \citep{DGK2007,PV2014,Bambi2021}. 

The most popular model for the hot medium is the inner hot geometrically thick optically thin flow \citep[e.g.][]{SL76,Ichimaru77,PKR97,Esin98,YN14}. 
It is supported by a number of arguments, including correlations between characteristic variability frequencies of the aperiodic noise \citep{ABL05}, the central frequencies of quasi-periodic oscillations  \citep{Pottschmidt03,Ibragimov05}, the spectral slope of Comptonization continuum, and the amplitude of Compton reflection \citep{Zdziarski99,Zdziarski03}.

A corona on top of a cold disk near the black hole was also considered as a source of hard X-rays \citep{HM93}. Simple models with a slab-like corona were found to produce too soft X-ray spectra \citep{Stern95}, because a large fraction of the coronal emission becomes reprocessed to soft radiation by the underlying disk, and the reflection component becomes stronger than observed. 
It was suggested that reprocessing could be reduced by the high ionization of the disk surface \citep{NayakshinDove01,Malzac05,PVZ18}. 

The observed spectrum may also be explained by a mildly relativistic outflowing corona, which beams hard X-rays away from the accretion disk, reducing their reflection and reprocessing \citep{Beloborodov99,Malzac01}. 
Outflows are generally expected to accompany the heating process in the magnetized corona. 
For instance, magnetic flares eject plasma, resembling solar flares.
Furthermore, if energy is released in compact flares, the plasma can become dominated by electron-positron pairs \citep{Svensson84,Stern95,PS96}, which have a tiny inertial mass. 
In this case, the plasma flows out with a saturated speed controlled by the local radiation field. 
Coronal outflows have also been observed in global simulations of accretion, which have begun to implement radiative effects in electron-ion plasma \citep{Liska22}.
Mildly relativistic ``winds'' are expected to surround the more relativistic ``jets'' seen in radio observations during the hard state \citep{Fender01,Stirling01}. 

In spite of a large body of the data on timing, spectra and imaging of BH X-ray binaries, there is still a large uncertainty in the geometry of their X-ray emission region.
Polarization has long been anticipated to provide an opportunity to determine the accretion geometry \citep{LS76}.
One of the best-studied BH X-ray binaries Cyg X-1 was observed with the {Imaging X-ray Polarimeter Explorer} (\ixpe)  \citep{Weisskopf2022} on 2022 May 15--21 and June 18--20. 
A rather high polarization degree PD=$4.0\pm0.2\%$ in the 2--8 keV range at the polarization angle PA=$-20\degr\pm2\degr$ consistent with the jet position angle \citep{Stirling01} was detected \citep{Krawczynski22}. 
Various models for the X-ray emitting region, such as a slab-corona, a static inner hot flow, and conical or spherical lamp-post corona at the disk rotation axis above the BH have been considered. 
\citet{Krawczynski22} suggest that the data can be reproduced only by models where the emission region is extended orthogonally to the jet (which is itself presumably perpendicular to the accretion disk). 
However, none of the considered models  reproduces the high PD at the measured inclination of $i=27.5^{+0.8}_{-0.6}$~deg \citep{2021Sci...371.1046M}; higher inclination, $i\gtrsim45\degr$ was required in order to reproduce the data.

In this paper, we show that the observed PD can be reproduced if the hot medium forms an outflow from the accretion disk. 
The mildly relativistic speed of the outflow affects polarization of scattered radiation, because the angular distribution of seed photons is significantly aberrated in the plasma rest frame \citep{Beloborodov98}. 
This results in a higher polarization of outgoing radiation at low inclinations. 
In Sect.~\ref{sec:model} we describe the setup of the model, and Sect.~\ref{sec:results} presents the results of our radiative transfer simulations. 
We discuss the obtained results and conclude in Sect.~\ref{sec:discussion}.  
The Appendices present the radiative transfer equation describing our model and a few examples of simulations. 

\section{Models for polarization} 
\label{sec:model} 

\subsection{Geometry} 
\label{sec:geom}

\begin{figure}
\centering
\includegraphics[width=0.95\columnwidth]{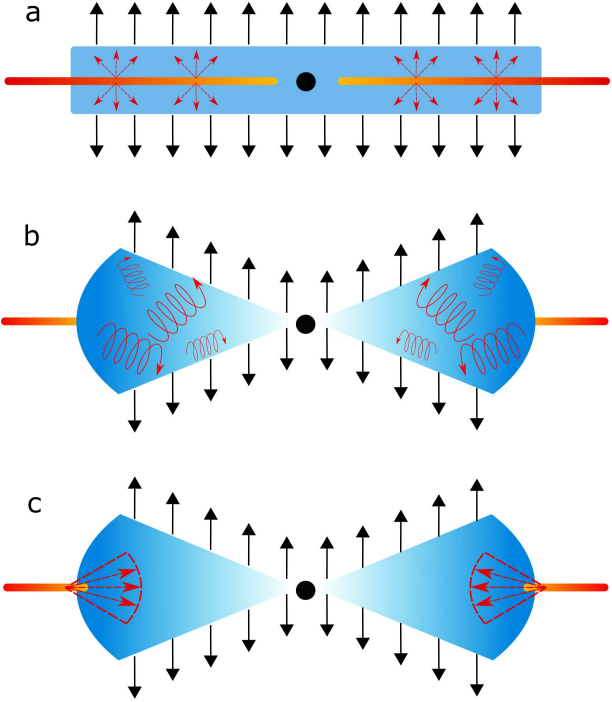}
\caption{Three alternative geometries considered in the work: 
panel (a) the outflowing corona with the underlying cold disk producing seed photons corresponding to model (A), panel (b) is for the outflowing hot flow with the internal synchrotron seed photons corresponding to model (B), and panel (c) is for the outflowing hot flow with the seed photons from the outer truncated disk (model C). } 
\label{fig:geometry}
\end{figure}

We assume that the observed X-rays in the hard state of Cyg X-1 are produced close to the compact object within the hot medium.  
We consider three alternative geometries sketched in Figure~\ref{fig:geometry}:  
\begin{enumerate}[(A)]
\item Hot corona covering the cold accretion disk (aka slab-corona) with the seed unpolarized blackbody photons of temperature $kT_{\rm bb}=0.1$~keV from the underlying accretion disk.  Radiation produced by multiple Compton scattering in such a slab is polarized at a level of $\sim$10\% \citep[e.g.][]{PS96,Schnittman10}.  
\item Hot medium situated within the truncated cold accretion disk; this model is usually referred to as a hot flow model.  
Here the seed photons could be produced within the flow, for example, by the synchrotron emission of non-thermal electrons with the peak at around 1~eV  \citep{PV09,Malzac09,VPV13}, or by cold clouds embedded in the hot flow \citep{Celotti92,PVZ18,Liska22}.  We consider the first case here. 
\item Same as  model (B), but with the seed unpolarized photons coming from the outer cold truncated disk of the temperature $kT_{\rm bb}=0.1$~keV.  
In this model, the incident angles of the seed photons are limited by the aspect ratio of the flow $h=H/R$ (see Appendix \ref{sec:appA} for details). 
\end{enumerate}
For all models, the scattering geometry is approximated by a plane-parallel slab extended along the disk plane. 

\subsection{Parameters of the outflow} 
\label{sec:outflow}

The hot medium itself can be in a dynamical state, where in addition to the azimuthal and radial motions, some fraction of the accreting matter leaves the system in a mildly relativistic wind. 
For our simulations, to account for the influence of the wind on polarization properties, we assume that the whole hot medium has the vertical velocity (aka outflowing corona) following the profile  
\begin{equation} 
\label{eq:beta_tau}
\beta(\tau)=\beta_0 \frac{\tau}{\taut} ,  
\end{equation} 
where $\tau$ is the Thomson optical depth measured from the central plane and $\taut$ is the total optical depth through the corona. 
The flow diverges sideways at some height $H$ which leads to a rapid drop of density. 
This height then can be considered as the upper boundary of the outflow.

We note here that the continuity equation implies that $n_{\rm e}(z) \beta(z)=n_0\beta_0$ is uniform with $z$ in a steady outflow.
For the velocity profile given by Equation~\eqref{eq:beta_tau}, the optical depth then can be expressed as 
\begin{equation}
d\tau(z)=n_{\rm e}(z) \sigmat  dz 
= \sigmat n_0  \frac{\beta_0}{\beta(z)} dz
=\sigmat n_0  \frac{\tau_0}{\tau(z)} dz , 
\end{equation}  
resulting in 
\begin{equation}
\tau(z)  =  \tau_0\, (z/H)^{1/2} ,
\qquad \tau_0=2\sigmat n_0 H.
\end{equation}   
We thus get the electron density and velocity dependence on height,
\begin{equation} 
n_{\rm e}(z)= n_0 (z/H)^{-1/2} , \qquad
\beta(z) = \beta_0 (z/H)^{1/2} .
\end{equation} 
The densest part of the outflow near the mid-plane has a small optical depth $\tau(z)\propto \sqrt{z}$ and makes a negligible contribution to scattering, so the details of how the inflow becomes the outflow near the mid-plane are not very important.

Our model implies mass loss rate from the disk of radius $R$ of   
\begin{equation}
\dot{M}_{\rm out}= 2 \pi R^2 c \beta_0  n_0 m_{\rm p} \mu_{\rm e}, 
\end{equation}   
where $\mu_{\rm e}$ is the mean molecular weight per electron in the outflow. On the other hand, the standard estimate for accretion rate in a disk with luminosity $L$ and radiative efficiency $\epsilon$ is 
\begin{equation}
\dot{M}_{\rm acc}= \frac{L}{\epsilon c^2} =  
\frac{\ell}{\epsilon} \frac{4 \pi R_{\rm g} m_{\rm p} c}{\sigmat} ,
\end{equation}   
where $\ell=L/L_{\rm Edd}$, $L_{\rm Edd}=4 \pi G M m_{\rm p} c /\sigmat$ is the Eddington luminosity, and $R_{\rm g}=GM/c^2$ is the gravitational radius.  
This gives
\begin{equation} 
\frac{\dot{M}_{\rm out}}{\dot{M}_{\rm acc}}
=  \frac{\epsilon R^2 \mu_{\rm e} }{\ell} \frac{\sigmat   \beta_0  n_0} {2R_{\rm g} } 
=  \frac{\epsilon \mu_{\rm e}\beta_0 \tau_0}{4\ell h}\,\frac{R}{R_{\rm g}}.
\end{equation}  
Our fiducial model has $h=1$, $\beta_0=1/2$, and $\taut=1$.  The observed luminosity of Cyg~X-1 implies $\ell\sim 0.02$, assuming the black hole mass $M\sim 10M_\odot$. Then, for an outflow made of electron-proton plasma ($\mu_{\rm e}=1$), we find $\dot{M}_{\rm out}/\dot{M}_{\rm acc}\sim 6 \epsilon R/R_{\rm g}$. This estimate is consistent with a large fraction of the accretion flow being diverted into an outflow before reaching the black hole. A significantly lower estimate for $\dot{M}_{\rm out}$ is found if the coronal heating occurs intermittently in space and time, in magnetic flares. In this case, the local radiation density can reach values sufficient for copious $e^\pm$ pair production, reducing $\mu_{\rm e}$ and $\dot{M}_{\rm out}$.

\subsection{Method} 
\label{sec:method}
 
The parameters of the corona are the electron temperature $T_{\rm e}$, the Thomson optical depth $\taut$, and the terminal velocity $\beta_0$ which is varied from 0 to 0.6. 
In the simulations we considered $kT_{\rm e}=100$~keV, which is close to the observed values \citep{Gierlinski1997}.
To reproduce the spectral energy distribution of Cyg~X-1 in the hard state \citep{Gierlinski1997,Krawczynski22}, we iterated $\taut$ to achieve the observed photon index $\Gamma=1.6$ in the 2--10~keV energy range. 

The radiative transfer equation that accounts for multiple Compton scattering of linearly polarized radiation in the moving medium is given in Appendix~\ref{sec:appA}.  
The only difference with the previously considered static models is that the effect of relativistic aberration has to be taken into account. 
We start simulations with some initial $\taut$ and solve radiative transfer equation in the plane-parallel (slab) approximation assuming azimuthal symmetry.   
Once the outgoing spectrum is obtained, we find the photon index $\Gamma$ in the 2--10~keV range (for an observer at $i= 30\degr$) and correct $\taut$. 
Iterations continue until the desired $\Gamma$ is reproduced. 
The escaping polarized radiation is described by two Stokes parameters $I$ and $Q$, with $U$ being identically zero because of azimuthal symmetry. 
We can then define PD$=Q/I$, which is positive when the dominant direction of oscillations of the electric vector (aka polarization vector) is parallel to the slab normal. 
If polarization vector is perpendicular to the flow normal, then PD is negative. 
For example, the optically thick electron-scattering atmosphere produces negative PD \citep{Cha60}.

The basic physics that controls the polarization production can be described as follows. 
Polarization of radiation scattered in the hot flow once depends strongly on the angular distribution of the incoming seed photons.
If they come from the underlying cold disk (and therefore beamed along the flow normal), the scattered radiation is polarized perpendicular to the flow normal (i.e. along the disk) resulting in negative PD. 
On the other hand, if they come from the outer truncated disk (i.e. sideways), the scattered radiation is polarized parallel to the normal, resulting in positive PD. 
In the latter case, polarization can reach 33\% in an ideal case when photons are injected in a plane and scattering is coherent  \citep{ST85,Poutanen22}.
If the corona is moving upwards, external radiation will be affected by relativistic aberration resulting in parallel polarization for $0.1\lesssim\beta\lesssim0.8$ \citep{Beloborodov98}.
When photons undergo many scatterings in an optically thin plane-parallel medium, polarization is in general parallel to the flow normal independently of the angular distribution of seed photons.

\begin{figure}
\centering
\includegraphics[width=0.85\columnwidth]{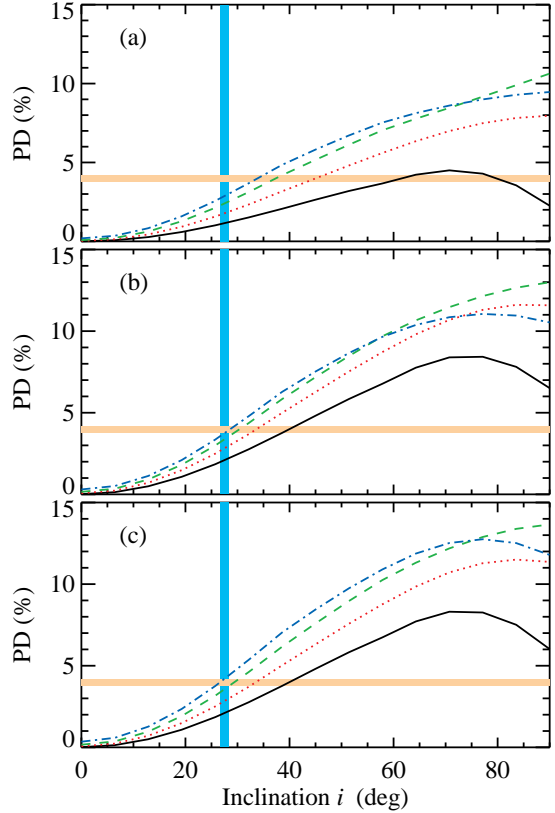}
\caption{Angular distribution of the PD for different outflow velocities  $\beta_0=0$ (black solid), 0.2 (red dotted), 0.4 (green dashed), and 0.6 (blue dot-dashed) in the middle of the IXPE range at 4~keV for the  three models are shown in Figure~\ref{fig:geometry}. 
Coronal parameters are  $kT_{\rm e}=100$ keV, $kT_{\rm bb}=0.1$ keV, $\Gamma=1.6$. 
The vertical blue stripe marks the $i=27.5^{+0.8}_{-0.6}$~deg inclination \citep{2021Sci...371.1046M} and the horizontal beige stripe corresponds to the observed X-ray PD from Cyg X-1 of $4.0\pm0.2\%$ \citep{Krawczynski22}. } 
\label{fig:angular}
\end{figure}

\section{Results} 
\label{sec:results}

Figure~\ref{fig:angular}a shows the angular distribution of the PD for model (A) in the middle of the IXPE range at 4~keV. 
We see that for static corona, the peak in PD is reached at $\sim$70\degr, while for the outflowing corona, the PD is growing with inclination more or less monotonically. 
This figure clearly shows that the static slab-corona can produce PD of 4\%  only at inclinations exceeding 60\degr. 
The outflowing corona, on the other hand, can produce this high PD at $i=30\degr$  for $\beta_0\gtrsim 0.6$. 

The angular distribution of the PD for model (B) is shown in  Figure~\ref{fig:angular}b. 
We see that the PD for the static flow of $\beta_0=0$ is larger than in the case of underlying disk seed photons. 
The reason is simple: the synchrotron seed photons take more scattering to get to the IXPE range resulting in a more anisotropic radiation field and higher PD. 
The observed 4\% PD is reached only at $i\approx 40\degr$.
For $\beta_0=0.2$, this PD is reached at inclination of $\approx 33\degr$, and the observed value of 4\% at $i<30\degr$ is produced when $\beta_0\gtrsim 0.4$.

The angular distribution of the PD for model (C) is shown in Figure~\ref{fig:angular}c. 
It is similar to the other models, but shows a slightly higher PD. 
We see that the PD observed in Cyg X-1 is well reproduced for the outflow velocity of $\beta_0\approx0.4$.  
We also considered a model of the seed photons from the cold clouds within the hot flow, but its results are very similar to those of models (b) and (c). 
We find that at all considered velocities models (b) and (c) predict parallel polarization exceeding $\sim$10\% at inclinations $i=$60\degr--70\degr.

\begin{figure}
\centering
\includegraphics[width=0.85\columnwidth]{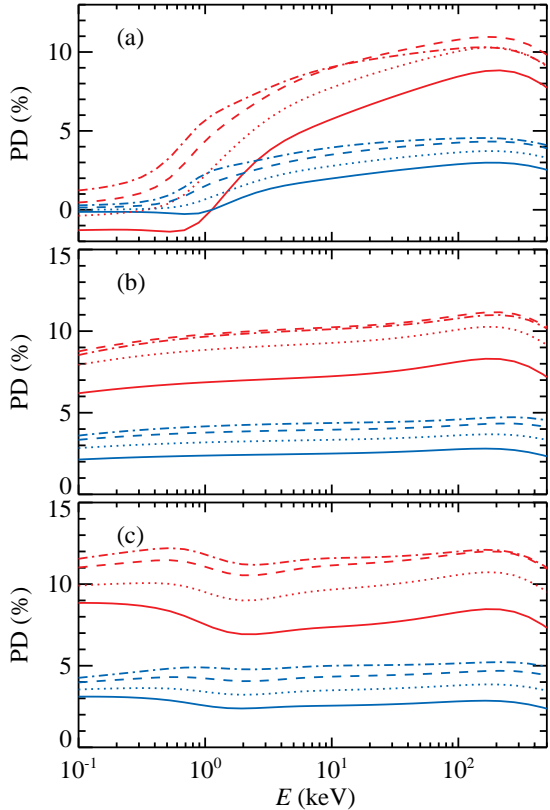}
\caption{Energy dependence of the polarization degree for the three models shown in Figure~\ref{fig:angular}. 
The blue and red lines correspond to the inclinations of  $30\degr$ and  $60\degr$, respectively. 
Solid, dotted, dashed and dot-dashed lines correspond to the outflow velocities of $\beta_0=0, 0.2, 0.4$ and 0.6, respectively. 
In the middle panel, the red curves for $\beta_0=0.4$ and 0.6 coincide. }
\label{fig:pd_energy}
\end{figure}

Let us now consider the energy dependence of PD. 
The results are shown in Figure~\ref{fig:pd_energy} for two selected inclinations of 30\degr\ and 60\degr.
For model (A) and the case of a static corona, polarization is negative (i.e. parallel to the disk) at photon energies where the first scattering dominates (see Appendix \ref{sec:appB} for spectral decomposition in different scattering orders), it changes the sign at around 1 keV and continues to grow to higher energies where the photons undergo more than two scatterings. 
At $i=30\degr$, PD does not exceed 3\% even at 100 keV.
We note that a static slab-corona model cannot produce hard spectra ($\Gamma\approx1.6$) observed in Cyg X-1 once the reflection and reprocessing of radiation is taken into account: thermalization of the hard radiation in the disk would produce too many soft photons that cool the corona producing soft spectra \citep{PVZ18}.

Motion of the corona away from the disk can be reconciled with the spectral hardness \citep{Beloborodov99} resulting also in a higher PD (see dotted, dashed and dot-dashed curves in Figure~\ref{fig:pd_energy}a). 
Because of the relativistic aberration, the seed photon angular distribution is less beamed along the normal in the flow comoving frame (see Figure~\ref{fig:slabcorona_details}), and even single-scattered photons have polarization parallel to the disc axis \citep[see also][]{Beloborodov98}. 
At an inclination $i<30\degr$, the PD of 4\% can be reached only for $\beta_0>0.6$, which also predicts an increasing trend of PD with energy.

Let us now consider a truncated disk geometry with the inner hot flow. 
Figure~\ref{fig:pd_energy}b shows the results for model (B) corresponding to the synchrotron seed photons. 
In the static case, the PD is 2--2.5\% at $i=30\degr$ (see blue solid line). 
Because photons reaching IXPE range undergo many scatterings, the PD is largely energy independent (see also Fig.~\ref{fig:ssc_details}). 
At the same inclination the PD increases with the outflow velocity reaching $\sim$4\% at $\beta_0\sim 0.4$ and 5\% at $\beta_0\sim 0.6$. 
At a higher inclination of $i=60\degr$, the PD is about twice as large as for the $i=30\degr$ case (see the set of red curves). 

Finally, we consider the model (C) with the hot flow geometry and the truncated disk providing seed photons for Comptonization.
The results for the case of $H/R=1$ are shown in Figure~\ref{fig:pd_energy}c.  
Here we see that the static hot flow produces nearly constant PD of $\sim$2.5\% in the IXPE range at $i=30\degr$. 
For the outflow velocity $\beta_0=0.2$, the anisotropy of the seed photons in the outflow rest frame is large enough to produce parallel polarization at a 4\% level for single-scattered photons, which then drops somewhat at higher scattering orders. 
At $\beta_0=0.4$, the PD reaches 4.5--5\% and weakly depends on energy in the IXPE range.
At even higher $\beta_0=0.6$, the PD reaches 6\%.

\section{Discussion and summary} 
\label{sec:discussion}

In our simulations, we ignored relativistic effects related to the rotation of the flow resulting in a rotation of the polarization plane due to relativistic effects. 
We note here that these effects are more important when the local emitted spectrum is a blackbody like, when relativistic aberration and Doppler boosting affect strongly the polarization vector just above the peak of the blackbody \citep{Connors1980,Dovciak2008,Loktev2022}.
However, for a power-law-like spectrum the effect is less pronounced because of a relatively larger contribution from large radii of the flow to the observed spectrum in a given energy range. 
Depolarization effects are $\lesssim1\%$ and can explain the difference between the required inclinations for the static models found in this work and in \citet{Krawczynski22}, where relativistic effects are taken into account.

Our simulations of the dynamic corona accounting for Comptonization confirm previous results for Thomson scattering that PD strongly depends on the angular distribution of incident photons \citep{Beloborodov98,BP99}. 
This effect is more important at the energies where the first scattering dominates. 
For the slab-corona, the static model results in a rapidly growing PD in the IXPE range, but it stays below 2\% for $i=30\degr$. 
Increasing velocity of the outflow gives a higher PD, which can reach the observed 4\% at $i=30\degr$ only when $\beta_0>0.6$. 
Such a high terminal velocity is required to produce the necessary photon anisotropy in the flow rest-frame due to relativistic aberration. 
We note that the average velocity of the outflow in that case is a factor of 2 smaller, and therefore is consistent with the estimate $\beta=0.3$ by \citet{Beloborodov99} that is required to achieve the photon spectral slope and the reflection fraction observed in Cyg X-1. 
The terminal velocity is somewhat larger than the equilibrium velocity of the electron-positron pairs, which may indicate the presence of a significant amount of protons and ions in the outflow and the role of magnetic processes (rather than radiation pressure) in the flow acceleration. 
 
The model with the internal synchrotron seed photons can reproduce  an energy-independent 4\% PD at $i\approx 30\degr$ for a more modest velocity of $\beta_0=$0.4--0.5 because in the IXPE range multiple scatterings are important and the angular distribution of seed photons does not play any role. 
Once we combine the mildly relativistic velocity with the anisotropy of the seed photons coming from the outer truncated disk, even higher PD can be achieved. 
We conclude that in order to get 4\% parallel polarization throughout the IXPE range assuming inclination of 30\degr, a mildly relativistic outflow velocity $\beta_0\approx0.4$ is needed together with either anisotropic distribution of seed soft photons coming from the outer cool disk or dominance of multiple scattering, which is a natural outcome of synchrotron photon injection at very lower energies.  

We also note that our model predicts PD exceeding 10\% at high inclinations $i>60\degr$, which potentially could be observed in other BH X-ray binaries.   
Similarly high polarization is also expected from highly inclined accretion flows in Seyfert galaxies of intermediate types \citep[see, e.g.,][for the case of NGC~4151]{Gianolli23}. 

\begin{acknowledgments}
We thank P.~Abolmasov and the referee A.~A.~Zdziarski for valuable suggestions. 
J.P. and A.V. acknowledge support from the Academy of Finland grant 333112.
A.M.B. is supported by NSF grant AST2009453, NASA grant 21-ATP21-0056, and Simons Foundation grant 446228. 
Nordita is supported in part by NordForsk.
\end{acknowledgments}

%

\vspace{5mm}
\facilities{\ixpe}


\software{ \sc{compps} \citep{PS96}}




\appendix

\section{Radiative transfer equation}
\label{sec:appA}

The radiative transfer equation (RTE) describing  Comptonization of polarized radiation in the plane-parallel atmosphere in the lab frame (marked with superscript ${\rm l}$) can be written in the form \citep{Mihalas84,NP94,BP99}: 
\begin{equation} 
\label{eq:RTE}
\mu \frac{d \vec{I}^{\rm l}(\tau,x,\mu)}{d \tau} = 
[1-\beta(\tau)\mu] \left[ - \sigma(x_{\rm c})  \vec{I}^{\rm l}(\tau,x,\mu) + \vec{S}^{\rm l}(\tau,x,\mu) \right]. 
\end{equation}
Here the photon energy is measured in the units of the electron rest mass $x=E/m_{\rm e}c^2$, $\mu$ is the cosine of the angle the photon momentum makes with the slab normal, $\vec{I}=(I,Q)^{\rm T}$ is the Stokes vector that fully describes linear polarization (Stokes $U$ parameter is zero due to the azimuthal symmetry), $\vec{S}^{\rm l}$ is the source function (also Stokes vector) in the lab frame, and  $\sigma(x_{\rm c})$ is the dimensionless total Compton scattering cross-section (in units of the Thomson cross-section $\sigmat$) for isotropic Maxwellian electron gas as a function of the photon energy in the comoving frame $x_{\rm c}= x/{\cal D}$, where ${\cal D}=1/[\gamma(1-\beta\mu)]$ is the Doppler factor, and $\gamma=1/\sqrt{1-\beta^2}$ is the Lorentz factor of the flow.  
Comoving frame quantities are marked with sub- or superscript ${\rm c}$.
We define the Thomson optical depth across the slab as measure of the column density:  $d\tau=n_{\rm e}(z)\sigmat\,dz$, where $z$ is the vertical coordinate and $n_{\rm e}$ is the electron concentration measured in the lab frame. 
The source function in the lab frame is related to that in the comoving frame by the Lorentz transformation \citep{RL79}
\begin{equation} \label{eq:sf_lor}
\vec{S}^{\rm l}(\tau,x,\mu) = {\cal D}^{3}\ \vec{S}^{\rm c}(\tau,x_{\rm c},\mu_{\rm c}) ,
\end{equation}
where the angles are related by the aberration formulae 
\begin{equation}\label{eq:aber}
\mu_{\rm c} = \frac{\mu-\beta}{1-\beta\mu} , \qquad 
\mu = \frac{\mu_{\rm c}+\beta}{1+\beta \mu_{\rm c}} . 
\end{equation}
The source function in the comoving frame is
\begin{equation} \label{eq:cs_source}
\vec{S}^{\rm c}(\tau,x_{\rm c},\mu_{\rm c})  
= x_{\rm c}^{2} \int_0^{\infty}\! \frac{\rmd x'_{\rm c}}{{x'_{\rm c}}^{2}} \int_{-1}^1\!\!\! \rmd \mu'_{\rm c} \ 
\mathbf{R} (x,\mu; x'_{\rm c},\mu'_{\rm c})  
\vec{I}^{\rm c}(\tau,x'_{\rm c},\mu'_{\rm c}) 
= {\cal R} \vec{I}^{\rm c},
\end{equation}
where $\mathbf{R}$ is the $2\times2$ azimuth-averaged redistribution matrix describing Compton scattering by isotropic hot electrons \citep[see Appendix A1 in][]{PS96}.  
The Stokes vectors $\vec{I}^{\rm c}$ in the comoving frame is related to that in the lab frame as 
\begin{equation} \label{eq:int_lor}
\vec{I}^{\rm c}(\tau,x_{\rm c},\mu_{\rm c})  = {\cal D}^{-3} \  \vec{I}^{\rm l}(\tau,x,\mu) .
\end{equation}

\begin{figure}
\centering
\includegraphics[width=0.8\columnwidth]{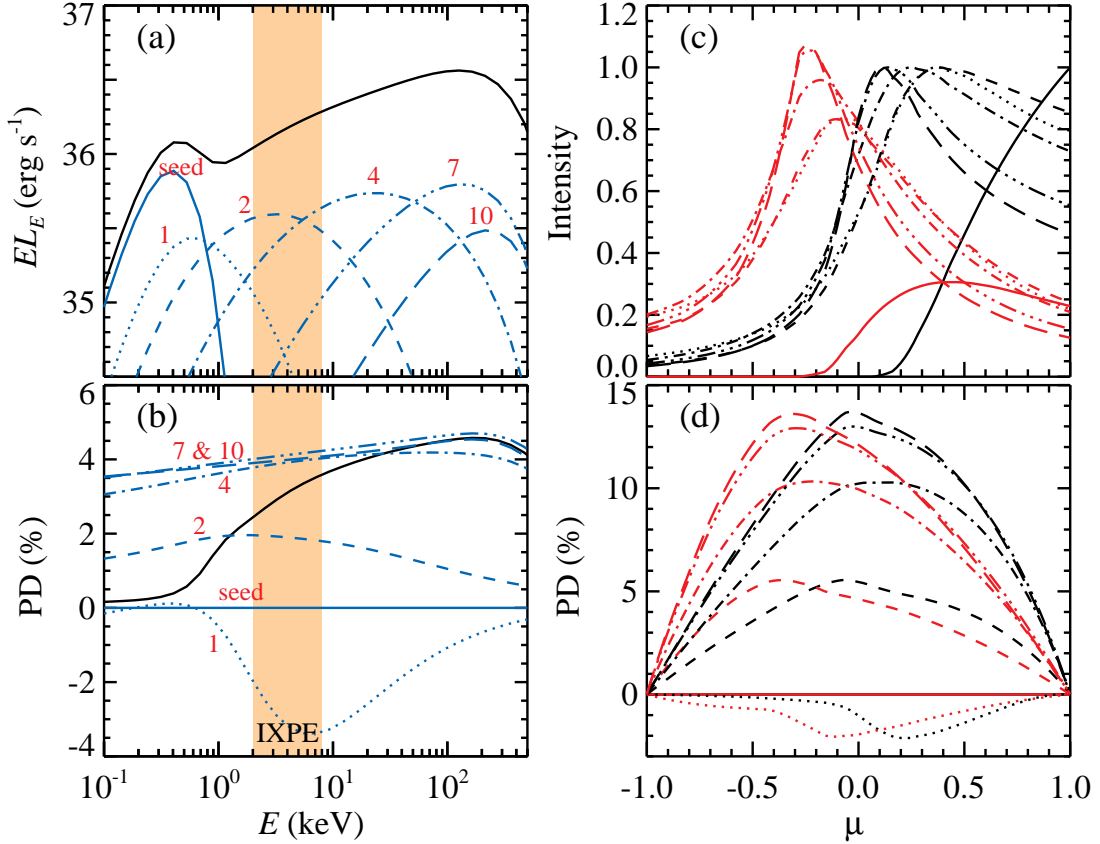}
\caption{Spectro-polarimetric properties of model (A) of a slab-corona of $T_{\rm e}=100$~keV and $T_{\rm bb}=0.1$~keV with the outflow of terminal velocity $\beta_0=0.4$. 
(a) Spectral energy distribution of the escaping radiation at inclination $i=30\degr$ scaled to produce the angle-integrated bolometric luminosity of $L=10^{37}$~erg~s$^{-1}$.
Blue solid, dotted, dashed, dot-dashed, triple-dot-dashed and long-dashed lines show the contribution of different scattering orders $n=0,1,2,4,7$ and 10, respectively. The black solid line gives the total spectrum. 
(b) The PD as a function of energy at the same inclination for the total radiation (black solid) and for  different scattering orders is shown. 
(c) Angular distribution of the intensity at optical depth $\tau=(3/4)\taut$ (where $\beta=0.3$) for the same scattering orders as above at the corresponding peaks of $EI_E$. The black and red curves correspond to the lab and comoving frame intensities, respectively.  
The lab-frame intensities are normalized to unity at the maximum. 
(d) Angular dependence of PD for the same scattering orders at the same energies as the intensity. 
The IXPE energy range is marked by a beige  vertical stripe in panels (a) and (b).}
\label{fig:slabcorona_details}
\end{figure}

\begin{figure}
\centering
\includegraphics[width=0.8\columnwidth]{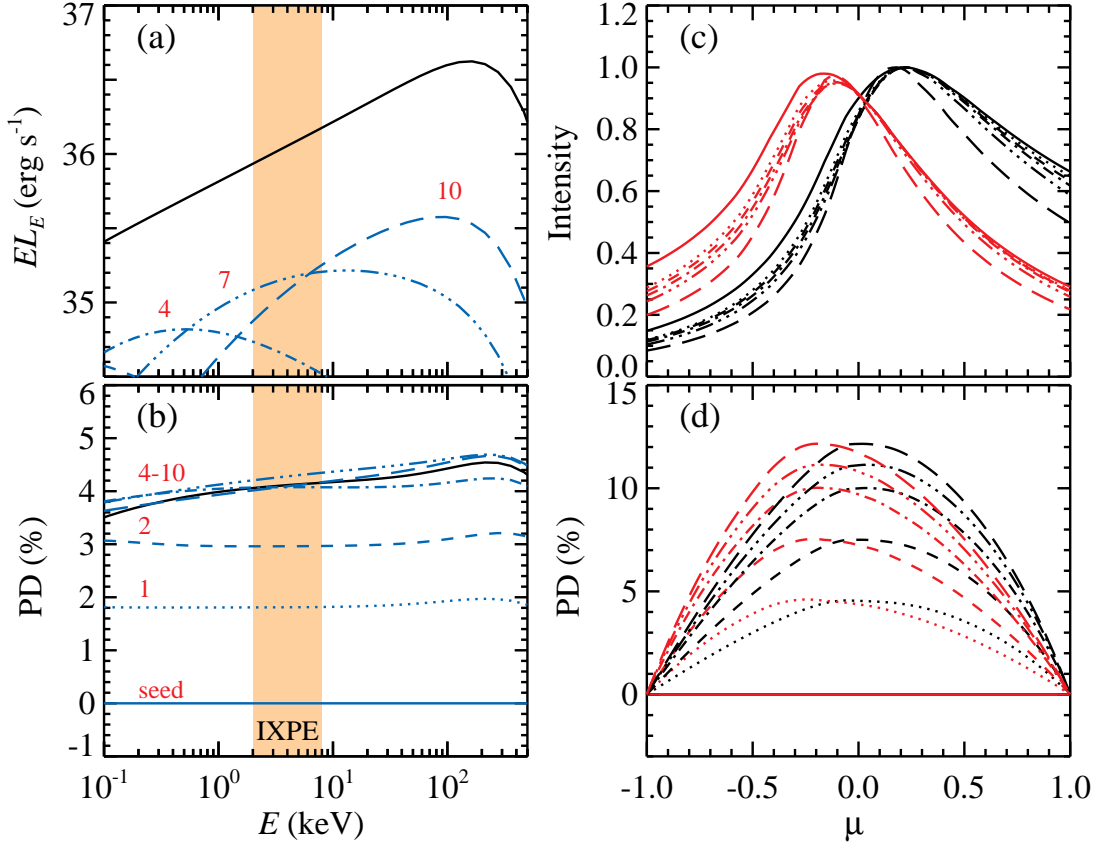}
\caption{Same as in Figure~\ref{fig:slabcorona_details}, but for the seed synchrotron photons of model (B).}
\label{fig:ssc_details}
\end{figure}

We solve the RTE using iterative scattering method \citep{PS96} accounting exactly for polarization by  Compton scattering on isotropic (in the flow frame) thermal electron gas \citep{NP93}. 
The only difference from \citet{PS96} is that here we account for bulk motion. 
At step 1, we specify the boundary condition at $\tau=0$, either on the bottom of the slab-corona, or in the middle of the slab, in the case of the hot flow--truncated disk geometry.
We approximate the spectrum of seed photons (marked by index 0) by the (unpolarized) blackbody 
\begin{equation}
\vec{I}^{\rm l}_0(\tau=0,x,\mu)=B_x(T_{\rm bb}) 
\left( \begin{array}{l}  
1 \\ 0 
\end{array} 
\right) {\cal H}(\mu_0-|\mu|), 
\end{equation}  
where ${\cal H}$ is the Heaviside step function, $\mu_0=1$ for the slab-corona, while for the truncated disk the injection of seed photons is limited to the zenith angles corresponding to the aspect ratio 
of the inner hot flow $h=H/R$, $|\mu|<\mu_0=h/\sqrt{1+h^2}$. 
At step 2, we use the formal solution of the RTE \eqref{eq:RTE} (with zero source function $\vec{S}$) to get the intensity in all other layers of the slab: 
\begin{equation}
\vec{I}^{\rm l}_0(\tau,x,\mu)= 
\vec{I}^{\rm l}_0(0,x,\mu)\ \exp(-\tau_x/\mu) , 
\end{equation}
where $\tau_x=\tau\sigma(x_{\rm c})[1-\overline{\beta}(\tau,0)\mu]$ is the energy- and angle-dependent optical depth measured from the bottom to a given layer and  $\overline{\beta}$ is the average  velocity from the slab center (or bottom for slab-corona model) $\tau=0$ to the given $\tau$, defined as 
\begin{equation}   \label{eq:beta_ave}
\overline{\beta}(\tau,\tau') = 
\frac{1}{\tau-\tau'} 
\int_{\tau'}^{\tau} \beta(t)\ dt = 
\frac{\beta(\tau)+\beta(\tau')}{2} = 
\beta_0 \frac{\tau+\tau'}{2\tau_0}.  
\end{equation}  
At step 3, we transform the Stokes vectors $\vec{I}^{\rm l}_n$ in the lab frame for photons scattered $n$ times to the comoving frame using Equation\,(\ref{eq:int_lor}).
At step 4, we compute the source function in the comoving frame for photons scattered $n+1$ times using Equation\,\eqref{eq:cs_source} as $\vec{S}^{\rm c}_{n+1} = {\cal R} \vec{I}^{\rm c}_n$.
At step 5, we get the source function in the lab frame using Equation\,\eqref{eq:sf_lor}.
And finally at step 6, we obtain the intensity of radiation scattered $n+1$ times from the formal solution of the RTE:
\begin{equation} \label{eq:formal_rte}
\vec{I}^{\rm l}_{n+1}(\tau,x,\mu)  = 
\left\{ 
\begin{array}{ll}  
\displaystyle 
\int_{\tau_{\min}}^{\tau}  \frac{\rmd \tau'}{\mu} \ \vec{S}^{\rm l}_{n+1}(\tau',x,\mu) \  [1-\beta(\tau')\mu]\ {\rm e}^{-(\tau-\tau')\sigma(x_{\rm c})[1-\overline{\beta}(\tau,\tau')\mu]/\mu}, & \mu>0, \\ 
& \\
\displaystyle \int_{\tau}^{\taut}  \frac{\rmd \tau'}{(-\mu)} \ \vec{S}^{\rm l}_{n+1}(\tau',x,\mu)\ [1-\beta(\tau')\mu] \ {\rm e}^{-(\tau'-\tau)\sigma(x_{\rm c})[1-\overline{\beta}(\tau',\tau)\mu]/(-\mu)}, & \mu<0,
\end{array}
\right.
\end{equation}   
where $\tau_{\min}=0$ for the slab-corona and $\tau_{\min}=-\taut$ for the hot flow, and $\overline{\beta}$ is given by Equation~\eqref{eq:beta_ave}. 
The iteration procedure  continues with steps 3--6 until the desired accuracy for the total Stokes vector $\vec{I}^{\rm l}=\sum_{n=0}^{\infty} \vec{I}^{\rm l}_n$ is achieved at all optical depths, angles and energies. 

For synchrotron seed photons, we start instead from step 4 and specify the source function in the comoving frame $\vec{S}^{\rm c}_0$. We assume it to be unpolarized, isotropic and optical-depth independent. 
Its spectral energy distribution is computed assuming a thermal electron distribution of $kT_{\rm e}=100$\,keV with a weak nonthermal tail. 
The resulting synchrotron spectrum has a shape of a broken power law with a peak at $\sim$1\,eV \citep{Wardzinski01,VPV13}.  
The iteration procedure now involves steps 5, 6 and then 3, 4, etc.

\section{Spectro-polarimetric properties of outflowing coronae}
\label{sec:appB}

We show a few examples of simulations of polarized radiation escaping from an outflowing corona for the same three models as considered in the main text. 
We concentrate here on the case $\beta_0=0.4$, which describes well Cyg X-1 data, and the observer inclination $i=30\degr$. 
Figure~\ref{fig:slabcorona_details} shows the main results for model (A), where seed unpolarized blackbody photons are injected from the bottom of outflowing corona. 
We see the spectral energy distribution of the total radiation as well as for selected scattering orders in panel (a).
The IXPE  2--8 keV range is dominated by photons scattered 2--4 times in the flow. 
Panel (b) shows the energy dependence of the PD. 
The single-scattered photons have small PD which becomes negative in the IXPE range, because the angular distribution of the seed photons in the comoving frame of the flow is still rather beamed outwards in spite of a high velocity as is seen in panel (c).  
The black curves there show the angular distribution of the intensity at the optical depth $\tau=(3/4)\taut$ at the energy, where $EI_E$ of the corresponding scattering order peaks.  
The red curves are the corresponding intensities in the comoving frame. 
We see that relativistic aberration is responsible for anisotropy of the radiation, with the intensity of multiple scattered photons in the lab frame peaking at $\mu\approx 0.1-0.3$ ($i\approx 70-80\degr$).
On the other hand,  the peak of the intensity in the comoving frame corresponds to $\mu_{\rm c}\approx -0.2$, i.e. it is directed more downwards. 
The angular distribution of the PD in both frames is shown in panel (d). 
We see that the PD grows with the scattering order. 
It reaches maximum roughly at the same angles where the intensity peaks. 
Because of the relativistic aberration, all curves in the lab-frame are shifted to higher $\mu$, so that at a given inclination, the PD grows with $\beta$.

\begin{figure}
\centering
\includegraphics[width=0.8\columnwidth]{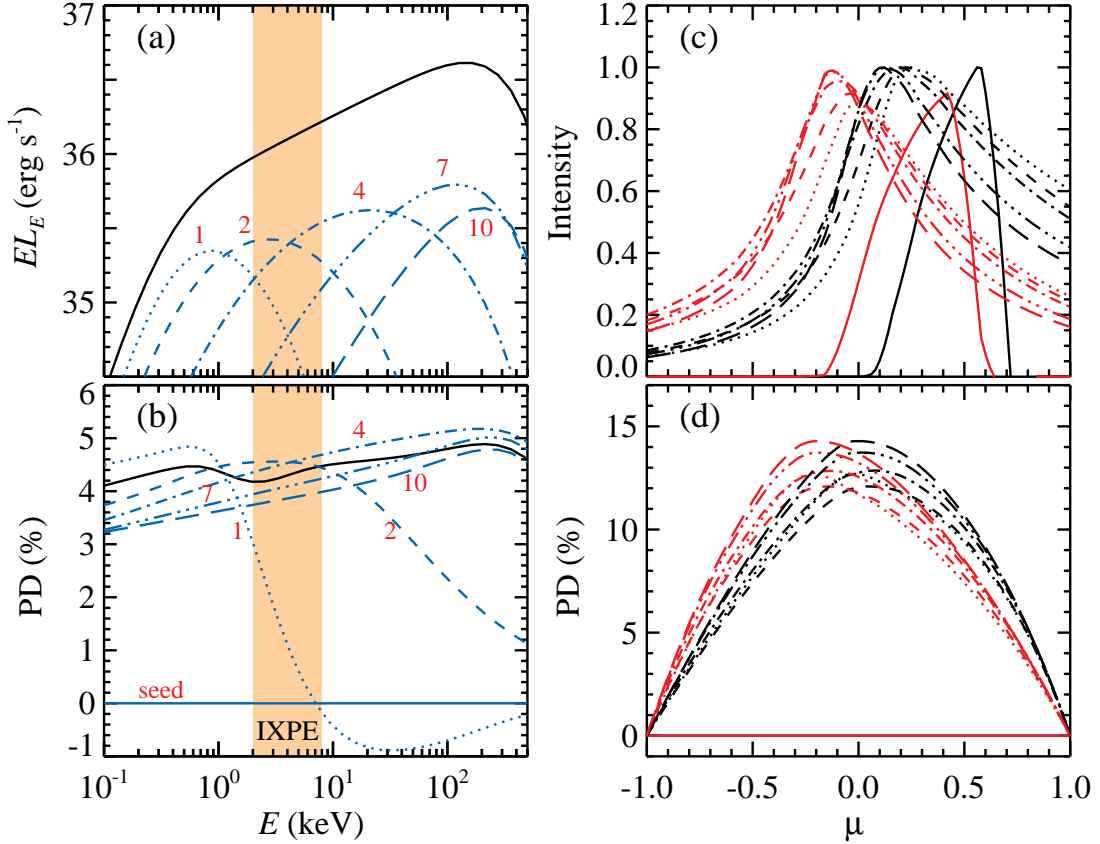}
\caption{Same as in Figure~\ref{fig:slabcorona_details}, but for the seed photons from the truncated disk with the coronal aspect ratio $H/R=1$ of model (C).}
\label{fig:trundisk_details}
\end{figure}

Figure~\ref{fig:ssc_details} shows the same quantities for model (B) with the seed synchrotron unpolarized photons which are isotropic in the comoving frame.  
The PD is increasing monotonically with the number of scatterings (panel d), but already at around $n>5$, very little variations is observed.  
Because about 7--8 scatterings is needed for seed photons to achieve the IXPE range, the PD is nearly energy-independent (panel b). 
The angular distributions of both the intensity and the PD (panels c and d) saturate. 

Figure~\ref{fig:trundisk_details} shows the results for model (C) with coronal aspect ratio $H/R=1$. 
In this model we see strong anisotropy of seed photons (panel c) both in the lab and comoving frames. 
This anisotropy results in a large parallel polarization reaching 12\% (for the cases of high inclination) already at the first scattering. 
The PD increases slightly at further scatterings having very similar behavior to the  previously considered models. 
The IXPE range is dominated by photons scattered 2--4 times as in the case of model (A), but now their PD is larger. 
The PD is nearly energy-independent as for the case of model (B), where photons scattered many more times contribute to the signal in the IXPE range.



\begin{thebibliography}{}
\expandafter\ifx\csname natexlab\endcsname\relax\def\natexlab#1{#1}\fi
\providecommand{\url}[1]{\href{#1}{#1}}
\providecommand{\dodoi}[1]{doi:~\href{http://doi.org/#1}{\nolinkurl{#1}}}
\providecommand{\doeprint}[1]{\href{http://ascl.net/#1}{\nolinkurl{http://ascl.net/#1}}}
\providecommand{\doarXiv}[1]{\href{https://arxiv.org/abs/#1}{\nolinkurl{https://arxiv.org/abs/#1}}}

\bibitem[{{Axelsson} {et~al.}(2005){Axelsson}, {Borgonovo}, \&
  {Larsson}}]{ABL05}
{Axelsson}, M., {Borgonovo}, L., \& {Larsson}, S. 2005, \aap, 438, 999,
  \dodoi{10.1051/0004-6361:20042362}

\bibitem[{{Bambi} {et~al.}(2021){Bambi}, {Brenneman}, {Dauser}, {Garc{\'\i}a},
  {Grinberg}, {Ingram}, {Jiang}, {Liu}, {Lohfink}, {Marinucci}, {Mastroserio},
  {Middei}, {Nampalliwar}, {Nied{\'z}wiecki}, {Steiner}, {Tripathi}, \&
  {Zdziarski}}]{Bambi2021}
{Bambi}, C., {Brenneman}, L.~W., {Dauser}, T., {et~al.} 2021, \ssr, 217, 65,
  \dodoi{10.1007/s11214-021-00841-8}

\bibitem[{{Beloborodov}(1998)}]{Beloborodov98}
{Beloborodov}, A.~M. 1998, \apjl, 496, L105, \dodoi{10.1086/311260}

\bibitem[{{Beloborodov}(1999)}]{Beloborodov99}
---. 1999, \apjl, 510, L123, \dodoi{10.1086/311810}

\bibitem[{{Beloborodov} \& {Poutanen}(1999)}]{BP99}
{Beloborodov}, A.~M., \& {Poutanen}, J. 1999, \apjl, 517, L77,
  \dodoi{10.1086/312051}

\bibitem[{{Celotti} {et~al.}(1992){Celotti}, {Fabian}, \& {Rees}}]{Celotti92}
{Celotti}, A., {Fabian}, A.~C., \& {Rees}, M.~J. 1992, \mnras, 255, 419,
  \dodoi{10.1093/mnras/255.3.419}

\bibitem[{{Chandrasekhar}(1960)}]{Cha60}
{Chandrasekhar}, S. 1960, {Radiative transfer} (New York: Dover)

\bibitem[{{Connors} {et~al.}(1980){Connors}, {Piran}, \& {Stark}}]{Connors1980}
{Connors}, P.~A., {Piran}, T., \& {Stark}, R.~F. 1980, \apj, 235, 224,
  \dodoi{10.1086/157627}

\bibitem[{{Done} {et~al.}(2007){Done}, {Gierli{\'n}ski}, \& {Kubota}}]{DGK2007}
{Done}, C., {Gierli{\'n}ski}, M., \& {Kubota}, A. 2007, \aapr, 15, 1,
  \dodoi{10.1007/s00159-007-0006-1}

\bibitem[{{Dov{\v{c}}iak} {et~al.}(2008){Dov{\v{c}}iak}, {Muleri}, {Goosmann},
  {Karas}, \& {Matt}}]{Dovciak2008}
{Dov{\v{c}}iak}, M., {Muleri}, F., {Goosmann}, R.~W., {Karas}, V., \& {Matt},
  G. 2008, \mnras, 391, 32, \dodoi{10.1111/j.1365-2966.2008.13872.x}

\bibitem[{{Esin} {et~al.}(1998){Esin}, {Narayan}, {Cui}, {Grove}, \&
  {Zhang}}]{Esin98}
{Esin}, A.~A., {Narayan}, R., {Cui}, W., {Grove}, J.~E., \& {Zhang}, S.-N.
  1998, \apj, 505, 854, \dodoi{10.1086/306186}

\bibitem[{{Fender}(2001)}]{Fender01}
{Fender}, R.~P. 2001, \mnras, 322, 31, \dodoi{10.1046/j.1365-8711.2001.04080.x}

\bibitem[{{Gianolli} {et~al.}(2023){Gianolli}, {Kim}, {Bianchi},
  {Ag{\'\i}s-Gonz{\'a}lez}, {Madejski}, {Marinucci}, {Matt}, {Middei},
  {Petrucci}, {Soffitta}, {Tagliacozzo}, {Tombesi}, {Ursini}, {Barnouin}, {De
  Rosa}, {Di Gesu}, {Ingram}, {Loktev}, {Panagiotou}, {Podgorny}, {Poutanen},
  {Puccetti}, {Ratheesh}, {Veledina}, {Zhang}, {Agudo}, {Antonelli},
  {Bachetti}, {Baldini}, {Baumgartner}, {Bellazzini}, {Bongiorno}, {Bonino},
  {Brez}, {Bucciantini}, {Capitanio}, {Castellano}, {Cavazzuti}, {Chen},
  {Ciprini}, {Costa}, {Del Monte}, {Di Lalla}, {Di Marco}, {Donnarumma},
  {Doroshenko}, {Dov{\v{c}}iak}, {Ehlert}, {Enoto}, {Evangelista}, {Fabiani},
  {Ferrazzoli}, {Garc{\'\i}a}, {Gunji}, {Heyl}, {Iwakiri}, {Jorstad}, {Kaaret},
  {Karas}, {Kislat}, {Kitaguchi}, {Kolodziejczak}, {Krawczynski}, {La Monaca},
  {Latronico}, {Liodakis}, {Maldera}, {Manfreda}, {Marscher}, {Marshall},
  {Massaro}, {Mitsuishi}, {Mizuno}, {Muleri}, {Negro}, {Ng}, {ODell}, {Omodei},
  {Oppedisano}, {Papitto}, {Pavlov}, {Peirson}, {Perri}, {Pesce-Rollins},
  {Pilia}, {Possenti}, {Ramsey}, {Rankin}, {Roberts}, {Romani}, {Sgr{\`o}},
  {Slane}, {Spandre}, {Swartz}, {Tamagawa}, {Tavecchio}, {Taverna}, {Tawara},
  {Tennant}, {Thomas}, {Trois}, {Tsygankov}, {Turolla}, {Vink}, {Weisskopf},
  {Wu}, {Xie}, \& {Zane}}]{Gianolli23}
{Gianolli}, V.~E., {Kim}, D.~E., {Bianchi}, S., {et~al.} 2023, \mnras,
  submitted, arXiv:2303.12541, \dodoi{10.48550/arXiv.2303.12541}

\bibitem[{{Gierlinski} {et~al.}(1997){Gierlinski}, {Zdziarski}, {Done},
  {Johnson}, {Ebisawa}, {Ueda}, {Haardt}, \& {Phlips}}]{Gierlinski1997}
{Gierlinski}, M., {Zdziarski}, A.~A., {Done}, C., {et~al.} 1997, \mnras, 288,
  958

\bibitem[{{Gierli{\'n}ski} {et~al.}(1999){Gierli{\'n}ski}, {Zdziarski},
  {Poutanen}, {Coppi}, {Ebisawa}, \& {Johnson}}]{Gierlinski1999}
{Gierli{\'n}ski}, M., {Zdziarski}, A.~A., {Poutanen}, J., {et~al.} 1999,
  \mnras, 309, 496, \dodoi{10.1046/j.1365-8711.1999.02875.x}

\bibitem[{{Haardt} \& {Maraschi}(1993)}]{HM93}
{Haardt}, F., \& {Maraschi}, L. 1993, \apj, 413, 507, \dodoi{10.1086/173020}

\bibitem[{{Ibragimov} {et~al.}(2005){Ibragimov}, {Poutanen}, {Gilfanov},
  {Zdziarski}, \& {Shrader}}]{Ibragimov05}
{Ibragimov}, A., {Poutanen}, J., {Gilfanov}, M., {Zdziarski}, A.~A., \&
  {Shrader}, C.~R. 2005, \mnras, 362, 1435,
  \dodoi{10.1111/j.1365-2966.2005.09415.x}

\bibitem[{{Ichimaru}(1977)}]{Ichimaru77}
{Ichimaru}, S. 1977, \apj, 214, 840, \dodoi{10.1086/155314}

\bibitem[{{Krawczynski} {et~al.}(2022){Krawczynski}, {Muleri}, {Dov{\v{c}}iak},
  {Veledina}, {Rodriguez Cavero}, {Svoboda}, {Ingram}, {Matt}, {Garcia},
  {Loktev}, {Negro}, {Poutanen}, {Kitaguchi}, {Podgorn{\'y}}, {Rankin},
  {Zhang}, {Berdyugin}, {Berdyugina}, {Bianchi}, {Blinov}, {Capitanio}, {Di
  Lalla}, {Draghis}, {Fabiani}, {Kagitani}, {Kravtsov}, {Kiehlmann},
  {Latronico}, {Lutovinov}, {Mandarakas}, {Marin}, {Marinucci}, {Miller},
  {Mizuno}, {Molkov}, {Omodei}, {Petrucci}, {Ratheesh}, {Sakanoi}, {Semena},
  {Skalidis}, {Soffitta}, {Tennant}, {Thalhammer}, {Tombesi}, {Weisskopf},
  {Wilms}, {Zhang}, {Agudo}, {Antonelli}, {Bachetti}, {Baldini}, {Baumgartner},
  {Bellazzini}, {Bongiorno}, {Bonino}, {Brez}, {Bucciantini}, {Castellano},
  {Cavazzuti}, {Ciprini}, {Costa}, {De Rosa}, {Del Monte}, {Di Gesu}, {Di
  Marco}, {Donnarumma}, {Doroshenko}, {Ehlert}, {Enoto}, {Evangelista},
  {Ferrazzoli}, {Gunji}, {Hayashida}, {Heyl}, {Iwakiri}, {Jorstad}, {Karas},
  {Kolodziejczak}, {La Monaca}, {Liodakis}, {Maldera}, {Manfreda}, {Marscher},
  {Marshall}, {Massaro}, {Mitsuishi}, {Ng}, {O'Dell}, {Oppedisano}, {Papitto},
  {Pavlov}, {Peirson}, {Perri}, {Pesce-Rollins}, {Pilia}, {Possenti},
  {Puccetti}, {Ramsey}, {Romani}, {Sgr{\`o}}, {Slane}, {Spandre}, {Tamagawa},
  {Tavecchio}, {Taverna}, {Tawara}, {Thomas}, {Trois}, {Tsygankov}, {Turolla},
  {Vink}, {Wu}, {Xie}, \& {Zane}}]{Krawczynski22}
{Krawczynski}, H., {Muleri}, F., {Dov{\v{c}}iak}, M., {et~al.} 2022, Science,
  378, 650, \dodoi{10.1126/science.add5399}

\bibitem[{{Lightman} \& {Shapiro}(1976)}]{LS76}
{Lightman}, A.~P., \& {Shapiro}, S.~L. 1976, \apj, 203, 701,
  \dodoi{10.1086/154131}

\bibitem[{{Liska} {et~al.}(2022){Liska}, {Musoke}, {Tchekhovskoy}, {Porth}, \&
  {Beloborodov}}]{Liska22}
{Liska}, M.~T.~P., {Musoke}, G., {Tchekhovskoy}, A., {Porth}, O., \&
  {Beloborodov}, A.~M. 2022, \apjl, 935, L1, \dodoi{10.3847/2041-8213/ac84db}

\bibitem[{{Loktev} {et~al.}(2022){Loktev}, {Veledina}, \&
  {Poutanen}}]{Loktev2022}
{Loktev}, V., {Veledina}, A., \& {Poutanen}, J. 2022, \aap, 660, A25,
  \dodoi{10.1051/0004-6361/202142360}

\bibitem[{{Malzac} \& {Belmont}(2009)}]{Malzac09}
{Malzac}, J., \& {Belmont}, R. 2009, \mnras, 392, 570,
  \dodoi{10.1111/j.1365-2966.2008.14142.x}

\bibitem[{{Malzac} {et~al.}(2001){Malzac}, {Beloborodov}, \&
  {Poutanen}}]{Malzac01}
{Malzac}, J., {Beloborodov}, A.~M., \& {Poutanen}, J. 2001, \mnras, 326, 417,
  \dodoi{10.1046/j.1365-8711.2001.04450.x}

\bibitem[{{Malzac} {et~al.}(2005){Malzac}, {Dumont}, \& {Mouchet}}]{Malzac05}
{Malzac}, J., {Dumont}, A.~M., \& {Mouchet}, M. 2005, \aap, 430, 761,
  \dodoi{10.1051/0004-6361:20041473}

\bibitem[{{McConnell} {et~al.}(2002){McConnell}, {Zdziarski}, {Bennett},
  {Bloemen}, {Collmar}, {Hermsen}, {Kuiper}, {Paciesas}, {Phlips}, {Poutanen},
  {Ryan}, {Sch{\"o}nfelder}, {Steinle}, \& {Strong}}]{McConnell02}
{McConnell}, M.~L., {Zdziarski}, A.~A., {Bennett}, K., {et~al.} 2002, \apj,
  572, 984, \dodoi{10.1086/340436}

\bibitem[{{Mihalas} \& {Mihalas}(1984)}]{Mihalas84}
{Mihalas}, D., \& {Mihalas}, B.~W. 1984, {Foundations of radiation
  hydrodynamics} (Mineola, NY: Dover)

\bibitem[{{Miller-Jones} {et~al.}(2021){Miller-Jones}, {Bahramian}, {Orosz},
  {Mandel}, {Gou}, {Maccarone}, {Neijssel}, {Zhao}, {Zi{\'o}{\l}kowski},
  {Reid}, {Uttley}, {Zheng}, {Byun}, {Dodson}, {Grinberg}, {Jung}, {Kim},
  {Marcote}, {Markoff}, {Rioja}, {Rushton}, {Russell}, {Sivakoff}, {Tetarenko},
  {Tudose}, \& {Wilms}}]{2021Sci...371.1046M}
{Miller-Jones}, J. C.~A., {Bahramian}, A., {Orosz}, J.~A., {et~al.} 2021,
  Science, 371, 1046, \dodoi{10.1126/science.abb3363}

\bibitem[{{Nagirner} \& {Poutanen}(1993)}]{NP93}
{Nagirner}, D.~I., \& {Poutanen}, J. 1993, \aap, 275, 325

\bibitem[{{Nagirner} \& {Poutanen}(1994)}]{NP94}
---. 1994, Astroph. Space Phys. Rev., 9, 1

\bibitem[{{Nayakshin} \& {Dove}(2001)}]{NayakshinDove01}
{Nayakshin}, S., \& {Dove}, J.~B. 2001, \apj, 560, 885, \dodoi{10.1086/323045}

\bibitem[{{Novikov} \& {Thorne}(1973)}]{NT73}
{Novikov}, D.~I., \& {Thorne}, K.~S. 1973, in Witt B., Witt C., eds, Les Astres
  Occlus (New York: Gordon \& Breach), 343

\bibitem[{{Pottschmidt} {et~al.}(2003){Pottschmidt}, {Wilms}, {Nowak},
  {Pooley}, {Gleissner}, {Heindl}, {Smith}, {Remillard}, \&
  {Staubert}}]{Pottschmidt03}
{Pottschmidt}, K., {Wilms}, J., {Nowak}, M.~A., {et~al.} 2003, \aap, 407, 1039,
  \dodoi{10.1051/0004-6361:20030906}

\bibitem[{{Poutanen} \& {Coppi}(1998)}]{PC98}
{Poutanen}, J., \& {Coppi}, P.~S. 1998, Physica Scripta, T77, 57.
\newblock \doarXiv{astro-ph/9711316}

\bibitem[{{Poutanen} {et~al.}(1997){Poutanen}, {Krolik}, \& {Ryde}}]{PKR97}
{Poutanen}, J., {Krolik}, J.~H., \& {Ryde}, F. 1997, \mnras, 292, L21

\bibitem[{{Poutanen} \& {Svensson}(1996)}]{PS96}
{Poutanen}, J., \& {Svensson}, R. 1996, \apj, 470, 249, \dodoi{10.1086/177865}

\bibitem[{{Poutanen} \& {Veledina}(2014)}]{PV2014}
{Poutanen}, J., \& {Veledina}, A. 2014, \ssr, 183, 61,
  \dodoi{10.1007/s11214-013-0033-3}

\bibitem[{{Poutanen} {et~al.}(2018){Poutanen}, {Veledina}, \&
  {Zdziarski}}]{PVZ18}
{Poutanen}, J., {Veledina}, A., \& {Zdziarski}, A.~A. 2018, \aap, 614, A79,
  \dodoi{10.1051/0004-6361/201732345}

\bibitem[{{Poutanen} \& {Vurm}(2009)}]{PV09}
{Poutanen}, J., \& {Vurm}, I. 2009, \apjl, 690, L97,
  \dodoi{10.1088/0004-637X/690/2/L97}

\bibitem[{{Poutanen} {et~al.}(2022){Poutanen}, {Veledina}, {Berdyugin},
  {Berdyugina}, {Jermak}, {Jonker}, {Kajava}, {Kosenkov}, {Kravtsov},
  {Piirola}, {Shrestha}, {Perez Torres}, \& {Tsygankov}}]{Poutanen22}
{Poutanen}, J., {Veledina}, A., {Berdyugin}, A.~V., {et~al.} 2022, Science,
  375, 874, \dodoi{10.1126/science.abl4679}

\bibitem[{{Remillard} \& {McClintock}(2006)}]{RM06}
{Remillard}, R.~A., \& {McClintock}, J.~E. 2006, \araa, 44, 49,
  \dodoi{10.1146/annurev.astro.44.051905.092532}

\bibitem[{{Rybicki} \& {Lightman}(1979)}]{RL79}
{Rybicki}, G.~B., \& {Lightman}, A.~P. 1979, {Radiative processes in
  astrophysics} (New York: Wiley)

\bibitem[{{Schnittman} \& {Krolik}(2010)}]{Schnittman10}
{Schnittman}, J.~D., \& {Krolik}, J.~H. 2010, \apj, 712, 908,
  \dodoi{10.1088/0004-637X/712/2/908}

\bibitem[{{Shakura} \& {Sunyaev}(1973)}]{1973A&A....24..337S}
{Shakura}, N.~I., \& {Sunyaev}, R.~A. 1973, \aap, 500, 33

\bibitem[{{Shapiro} {et~al.}(1976){Shapiro}, {Lightman}, \& {Eardley}}]{SL76}
{Shapiro}, S.~L., {Lightman}, A.~P., \& {Eardley}, D.~M. 1976, \apj, 204, 187,
  \dodoi{10.1086/154162}

\bibitem[{{Stern} {et~al.}(1995){Stern}, {Poutanen}, {Svensson}, {Sikora}, \&
  {Begelman}}]{Stern95}
{Stern}, B.~E., {Poutanen}, J., {Svensson}, R., {Sikora}, M., \& {Begelman},
  M.~C. 1995, \apjl, 449, L13, \dodoi{10.1086/309617}

\bibitem[{{Stirling} {et~al.}(2001){Stirling}, {Spencer}, {de la Force},
  {Garrett}, {Fender}, \& {Ogley}}]{Stirling01}
{Stirling}, A.~M., {Spencer}, R.~E., {de la Force}, C.~J., {et~al.} 2001,
  \mnras, 327, 1273, \dodoi{10.1046/j.1365-8711.2001.04821.x}

\bibitem[{{Sunyaev} \& {Titarchuk}(1985)}]{ST85}
{Sunyaev}, R.~A., \& {Titarchuk}, L.~G. 1985, \aap, 143, 374

\bibitem[{{Svensson}(1984)}]{Svensson84}
{Svensson}, R. 1984, \mnras, 209, 175, \dodoi{10.1093/mnras/209.2.175}

\bibitem[{{Veledina} {et~al.}(2013){Veledina}, {Poutanen}, \& {Vurm}}]{VPV13}
{Veledina}, A., {Poutanen}, J., \& {Vurm}, I. 2013, \mnras, 430, 3196,
  \dodoi{10.1093/mnras/stt124}

\bibitem[{{Wardzi{\'n}ski} \& {Zdziarski}(2001)}]{Wardzinski01}
{Wardzi{\'n}ski}, G., \& {Zdziarski}, A.~A. 2001, \mnras, 325, 963,
  \dodoi{10.1046/j.1365-8711.2001.04387.x}

\bibitem[{{Weisskopf} {et~al.}(2022){Weisskopf}, {Soffitta}, {Baldini},
  {Ramsey}, {O'Dell}, {Romani}, {Matt}, {Deininger}, {Baumgartner},
  {Bellazzini}, {Costa}, {Kolodziejczak}, {Latronico}, {Marshall}, {Muleri},
  {Bongiorno}, {Tennant}, {Bucciantini}, {Dovciak}, {Marin}, {Marscher},
  {Poutanen}, {Slane}, {Turolla}, {Kalinowski}, {Di Marco}, {Fabiani},
  {Minuti}, {La Monaca}, {Pinchera}, {Rankin}, {Sgro'}, {Trois}, {Xie},
  {Alexander}, {Allen}, {Amici}, {Andersen}, {Antonelli}, {Antoniak},
  {Attin{\`a}}, {Barbanera}, {Bachetti}, {Baggett}, {Bladt}, {Brez}, {Bonino},
  {Boree}, {Borotto}, {Breeding}, {Brienza}, {Bygott}, {Caporale}, {Cardelli},
  {Carpentiero}, {Castellano}, {Castronuovo}, {Cavalli}, {Cavazzuti},
  {Ceccanti}, {Centrone}, {Citraro}, {D'Amico}, {D'Alba}, {Di Gesu}, {Del
  Monte}, {Dietz}, {Di Lalla}, {Persio}, {Dolan}, {Donnarumma}, {Evangelista},
  {Ferrant}, {Ferrazzoli}, {Ferrie}, {Footdale}, {Forsyth}, {Foster},
  {Garelick}, {Gunji}, {Gurnee}, {Head}, {Hibbard}, {Johnson}, {Kelly},
  {Kilaru}, {Lefevre}, {Roy}, {Loffredo}, {Lorenzi}, {Lucchesi}, {Maddox},
  {Magazzu}, {Maldera}, {Manfreda}, {Mangraviti}, {Marengo}, {Marrocchesi},
  {Massaro}, {Mauger}, {McCracken}, {McEachen}, {Mize}, {Mereu}, {Mitchell},
  {Mitsuishi}, {Morbidini}, {Mosti}, {Nasimi}, {Negri}, {Negro}, {Nguyen},
  {Nitschke}, {Nuti}, {Onizuka}, {Oppedisano}, {Orsini}, {Osborne}, {Pacheco},
  {Paggi}, {Painter}, {Pavelitz}, {Pentz}, {Piazzolla}, {Perri},
  {Pesce-Rollins}, {Peterson}, {Pilia}, {Profeti}, {Puccetti}, {Ranganathan},
  {Ratheesh}, {Reedy}, {Root}, {Rubini}, {Ruswick}, {Sanchez}, {Sarra},
  {Santoli}, {Scalise}, {Sciortino}, {Schroeder}, {Seek}, {Sosdian}, {Spandre},
  {Speegle}, {Tamagawa}, {Tardiola}, {Tobia}, {Thomas}, {Valerie}, {Vimercati},
  {Walden}, {Weddendorf}, {Wedmore}, {Welch}, {Zanetti}, \&
  {Zanetti}}]{Weisskopf2022}
{Weisskopf}, M.~C., {Soffitta}, P., {Baldini}, L., {et~al.} 2022, JATIS, 8,
  026002, \dodoi{10.1117/1.JATIS.8.2.026002}

\bibitem[{{Yuan} \& {Narayan}(2014)}]{YN14}
{Yuan}, F., \& {Narayan}, R. 2014, \araa, 52, 529,
  \dodoi{10.1146/annurev-astro-082812-141003}

\bibitem[{{Zdziarski} \& {Gierli{\'n}ski}(2004)}]{ZG04}
{Zdziarski}, A.~A., \& {Gierli{\'n}ski}, M. 2004, Progr. Theor. Phys. Suppl.,
  155, 99, \dodoi{10.1143/PTPS.155.99}

\bibitem[{{Zdziarski} {et~al.}(2001){Zdziarski}, {Grove}, {Poutanen}, {Rao}, \&
  {Vadawale}}]{Zdziarski01}
{Zdziarski}, A.~A., {Grove}, J.~E., {Poutanen}, J., {Rao}, A.~R., \&
  {Vadawale}, S.~V. 2001, \apjl, 554, L45, \dodoi{10.1086/320932}

\bibitem[{{Zdziarski} {et~al.}(2003){Zdziarski}, {Lubi{\'n}ski}, {Gilfanov}, \&
  {Revnivtsev}}]{Zdziarski03}
{Zdziarski}, A.~A., {Lubi{\'n}ski}, P., {Gilfanov}, M., \& {Revnivtsev}, M.
  2003, \mnras, 342, 355, \dodoi{10.1046/j.1365-8711.2003.06556.x}

\bibitem[{{Zdziarski} {et~al.}(1999){Zdziarski}, {Lubi{\'n}ski}, \&
  {Smith}}]{Zdziarski99}
{Zdziarski}, A.~A., {Lubi{\'n}ski}, P., \& {Smith}, D.~A. 1999, \mnras, 303,
  L11, \dodoi{10.1046/j.1365-8711.1999.02343.x}

\end{thebibliography}



\end{document}